# Pseudo Dynamic Transitional Modeling of Building Heating Energy Demand Using Artificial Neural Network

Subodh Paudel[a,b,c], Mohamed Elmtiri[b], Wil L. Kling[c], Olivier Le Corre[a*], Bruno Lacarrière[a]

[a]Department of Energy System and Environment, Ecole des Mines, Nantes, GEPEA, CNRS, UMR 6144, France
[b]Environnement Recherche et Innovation, Veolia, France
[c]Department of Electrical Engineering, Technische Universiteit Eindhoven, Netherlands
*Corresponding author. Tel.: +33 2 51 85 82 57
E-mail: Olivier.Lecorre@mines-nantes.fr

**Abstract**

This paper presents the building heating demand prediction model with occupancy profile and operational heating power level characteristics in short time horizon (a couple of days) using artificial neural network. In addition, novel pseudo dynamic transitional model is introduced, which consider time dependent attributes of operational power level characteristics and its effect in the overall model performance is outlined. Pseudo dynamic model is applied to a case study of French Institution building and compared its results with static and other pseudo dynamic neural network models. The results show the coefficients of correlation in static and pseudo dynamic neural network model of 0.82 and 0.89 (with energy consumption error of 0.02%) during the learning phase, and 0.61 and 0.85 during the prediction phase respectively. Further, orthogonal array design is applied to the pseudo dynamic model to check the schedule of occupancy profile and operational heating power level characteristics. The results show the new schedule and provide the robust design for pseudo dynamic model. Due to prediction in short time horizon, it finds application for Energy Services Company (ESCOs) to manage the heating load for dynamic control of heat production system.

## 1. Introduction

The global concerns of climate change and regulation in energy emissions have drawn more attention towards researchers and industries for the design and implementation of energy systems for low energy buildings. According to IEA statistics [1], total energy use globally accounts for around 7200 Mtoe (Mega Tonnes Oil Equivalents). Residential and commercial buildings consume 40% of final energy use in the world and European countries consume 76% of energy towards thermal comfort in buildings. The small deviations in design parameters of buildings could bring large adverse effect in the energy efficiency and which, additionally, results in huge emissions from the buildings. It is estimated that improvement in energy efficiency of the buildings in European Union by 20% will result in saving at least 60 billion Euro annually [2]. So, research is very active in driving towards the sustainable/low energy buildings. In order to accomplish this and to ensure thermal comfort, it is essential to know energy flows and energy demand of the buildings for the control of heating and cooling energy production from plant systems. The energy demand of the building system, thus, depends on physical and geometrical parameters of buildings, operational characteristics of heating and cooling energy plant systems, weather conditions, appliances characteristics and internal gains.

There are various approaches to predict building energy demand based on physical methods and data-driven methods (statistical and regression methods and artificial intelligence methods) as mentioned by Zhao et al. [3]. Physical methods are based on physical engineering methods and uses thermodynamics and heat transfer characteristics to determine the energy demand of the building. There are numerous physical simulation tools developed as EnergyPlus [4], ESP-r [5], IBPT [6], SIMBAD [7], TRNSYS [8], CARNOT [9] etc… to compute the building energy demand. A simplified physical model based on physical, geometrical, climatic and occupant model was presented by

Duanmu et al. [10] to bridge the complexities of collecting more physical data required in simulation tools. Other possible approaches for building energy prediction are semi-physical models like response factor method, transfer function method, frequency analysis method and lumped method [11]. Though methodologies adapted to estimate energy demand of buildings are different in physical and semi-physical models, both are highly parameterized. In addition, physical parameters of buildings are not always known or even sometimes data are missing. And also, these models are computationally expensive for Energy Services Company (ESCOs) to manage heating and cooling loads for control applications.

Other approaches to predict building energy demand with limited physical parameters are data-driven methods, which strongly dependent on the measurements of historical data. Statistical and regression methods seem more feasible to predict building energy demand with limited physical parameters. The statistical approaches have been widely used by Girardin et al. [12] to determine the best model parameters by fitting actual data. Different approaches (physical and behaviour characteristics based on statistical data) were presented by Yao et al. [13] to bridge the gap between semi-physical and statistical methods. In their work, statistical daily load profile was grounded on energy consumption per capita and human behaviour factor, and semi-physical method was based on thermal resistance capacitance network. Nevertheless, these statistical models used linear characteristics of input and output variables to evaluate the building parameters and are not adapted to non-linear energy demand behavior. Regression models [14-15] have also been used to predict the energy demand, but, they are not accurate enough to represent short term horizon (couple of days) with hourly (or couple of minutes) sampling time energy demand prediction. In order to find the best fitting from the actual data, this kind of models requires significant effort and time.

In recent years, there is a growth in research work in the field of artificial intelligence (AI) like artificial neural network [3, 16] and support vector machines [3, 17-18]. These methods are known for solving the complex non-linear function of energy demand models with limited physical parameters. Neural network method has shown better performances than physical, statistical and regression methods. Authors [19-20] used static neural network to predict energy demand of the building and compared results with physical models. For instance, Kalogirou et al. [19] used climate variables (mean and maximum of solar radiation, wind speed, and other parameters as wall and roof type) coupled with artificial neural network (ANN) to predict daily heating and cooling load of the buildings. In their work, results obtained using ANN are similar to those given by the physical modelling tool TRNSYS. Neto et al. [20] presented a comparison of neural network approach with physical simulation tool EnergyPlus. In this work, authors used climate variables as external dry temperature, relative humidity and solar radiation as input variables to predict daily consumption of the building. Results showed that neural network is slightly more accurate than EnergyPlus when comparing with real data. Static neural network model proposed by Shilin et al. [21] consider climate variables as dry bulb temperature and information regarding schedule of holiday's to predict cooling power of residential buildings. Dong et al. [17] used support vector machine (SVM) to predict the monthly building energy consumption using dry bulb temperature, relative humidity and global solar radiation. Performance of SVM and neural network model wee compared and results show that SVM was better than neural network in prediction.

Various authors [22-26] performed hourly building energy prediction using ANN. Mihalakakou et al. [22] performed hourly prediction of residential buildings with solar radiation and multiple delays of air temperature predictions as input variables. Ekici et al. [23] used building parameters (window's transmittivity, building's orientation, and insulation thickness) and Dombayci [24] used time series information of hour, day and month, and energy consumption of the previous hour to predict the hourly heating energy consumptions. Gonzalez et al. [25] used time series information hour and day, current energy consumption and predicted values of temperature as input variables to predict hourly energy consumption of building system. Popescu et al. [26] used climate variables as solar radiation, wind speed, outside temperature of previous 24 hours, and other variables as mass flow rate of hot water of previous 24 hours and hot water temperature exit from plant system to predict the space hourly heat consumptions of buildings. Li et al. [18] used SVM to predict hourly cooling load of office building using climate variables as solar radiation, humidity and outdoor temperature. In their work, SVM was compared with static neural network and result showed SVM better than static neural network in terms of model performance. Dynamic neural network method which includes time dependence was presented by Kato et al. [27] to predict heating load of district heating and cooling system based on maximum and minimum air temperature. Kalogirou et al. [28] used Jordan Elman recurrent dynamic network to predict energy consumption of a passive solar building system based on seasonal information, masonry thickness and thermal insulation.

For many authors [29-31] occupancy profile has a significant impact on building energy consumption. Sun et al. [29] mentioned that occupancy profile period has a significant impact on initial temperature requirement in the building during morning. In their work, reference day (the targeted day prediction which depends on previous day and beginning of following day based on occupancy and non-occupancy profile period) was calculated based on occupancy profile period. In addition to this value, correlated weather data and prediction errors of previous 2 hours were used as input variables to predict hourly cooling load. Yun et al. [30] used ARX (autoregressive with exogeneous i.e., external, inputs) time and temperature indexed model with occupancy profile to predict hourly heating and cooling load of building system and compared this with results given by neural network. Results showed that occupancy profile has a significant contribution in determination of auto regressive terms during different intervals of time and further showed a variation of it in the building heating and cooling energy consumption. The proposed ARX model showed similar performance with neural network. Sensitivity analysis for heating, cooling, hot water, equipment and lighting energy consumption based on occupancy profile was performed by Azar et al. [31] for different sizes of office buildings. In their work, they found that heating energy consumption has the highest sensitivity compared to cooling, hot water, equipment and lighting energy consumption for small size buildings. Also, results showed that heating energy consumption is highly influenced by occupancy profile for medium and small buildings during the occupancy period. Moreover, few literatures focused on operational power level characteristics (schedule of heating and cooling energy to manage energy production from plant system). For example, Leung et al. [32] used climate variables and operational characteristics of electrical power demand (power information of lighting, air-conditioning and office equipment which implicitly depends on occupancy schedule of electrical power demand) to predict hourly and daily building cooling load using neural network.

In conclusion, it can be reiterated that physical and semi-physical models [4-11], though give precise prediction of building energy, they are highly parameterized and are computationally expensive to manage the energy for control applications for ESCOs. Data-driven methods which depend on measurement historical data are not effective during the early stage of building operation and construction since measurement data are not available at these stages. When building energy data are available, data-driven methods can be considered if measurement data are accurate and reliable as this kind of models can be sensitive on the quality of measured data. Sensitivity of the accuracy of data driven models, thus, depends on the measurement data. Data-driven models based on statistical and regression methods [12-15, 26] cannot precisely represent short time horizon (couple of days) with hourly (or couple of minutes) sampling time prediction, though they perform prediction of energy consumptions of buildings with limited physical parameters. They also require significant efforts and time to compute the best fitting of the actual data. Static neural network models [19-21] are used for daily prediction and [22-25] are used for hourly prediction of the buildings energy consumptions. Though dynamic neural network model [27-28] gives better precision in compared to static neural network, they do not consider occupancy profile and operational power level characteristics of the plant system and therefore not adapted for the ESCOs to manage energy production for control applications. The important features like transition and time dependent attributes of operational power level characteristics of the plant system are still missing, though, authors [29-30] consider occupancy profile and author [32] considers operational characteristics of electrical power demand. The detailed variables and application of models developed in the literature reviews are summarized in Table 1.

Table 1 : Summary of variables and application models in the literature

| Author and Year | Type of Model | Climate Variables | | | | | | | Occupancy Profile | Operational Characteristics | Other Parameters | Horizon of Forecast | Type of Applications for Buildings |
| | | Outside Tempeature | | | Inner Temperature | Global Solar Radiation | Wind Speed | Relative Humidity | | | | | |
| | | Ambient | Dry Bulb | Wet Bulb | | | | | | | | | |
|---|---|---|---|---|---|---|---|---|---|---|---|---|---|
| Girardin et al. (2009) | Statistical | √ | | | | | | | | | √ (1*) | Annually | 80 Residential (heating and cooling) |
| Yao et al. (2005) | Thermal and Statistical | √ | | | | | | | | | √ (2*) | Daily | Residential (space heating) |
| Catalina et al. (2008) | Regression | √ | | | | √ | | | | | | Monthly | Residential |
| Wan et al. (2012) | Regression | | √ | √ | | √ | | | | | √ (3*) | Monthly & Yearly | Office (heating and cooling) |
| Dong et al. (2005) | SVM | | √ | | | √ | | √ | | | | Monthly | 4 Buildings (total energy consumptions) |
| Kalogirou et al. (2001) | Static NN | | | | | √ | | √ | | | | Daily | 9 Buildings (heating and cooling) |
| Neto et al. (2008) | Static NN | | √ | | | √ | | √ | | | | Daily | Office (3000 m2) |
| Shilin et al. (2010) | Static NN | √ | | | | | | | | | | Daily | Residential (cooling power) |
| Mihalakakou et al. (2002) | Static NN | √(4*) | | | | √ | | | | | | Hourly | Residential (200 m2) |
| Ekici et al. (2009) | Static NN | | | | | | | | | | √ (5*) | Hourly | Heating Energy of Buildings |
| Dombayci (2010) | Static NN | | | | | | | | | | √ (6*) | Hourly | Residential (heating energy) |
| Gonzalez et al. (2005) | Static NN | √ | | | | | | | | | √ (7*) | Hourly | Electrical load |
| Popescu et al. (2009) | Static NN | | | | | √ | √ | √ | | | √ (8*) | Hourly | 8 Buildings |
| Kato et al. (2008) | Dynamic NN | √(9*) | | | | | | | | | | Hourly | District (heating energy) |
| Kalogirou et al. (2000) | Dynamic NN | | | | | | | | | | √ (10*) | Hourly | Passive solar buildings |
| Li et al. (2010) | SVM | | √(11*) | | | √(11*) | √ | | | | | Hourly | Office building and library |
| Sun et al. (2013) | Regression | √ | | | | √ | | √ | √(12*) | | √ (13*) | Hourly | Cooling load for high rise buildings (440,000 m2) |
| Yun et al. (2012) | Autoregressive with exogenous | √ | √ | | | √ | √ | √ | √ | | | Hourly | Small building for heating load (464 m2) |
| Leung et al. (2012) | Static NN | | √ | √ | | √ | √ | | √(14*) | √ | √ (15*) | Hourly & Daily | Office (space electrical power demand) |
| Duanmu et al. (2013) | Physical | √ | | | √ | √ | | √ | | | √ (16*) | Hourly | Cooling load of buildings |

Remarks:

1*: Nominal Temperature of heating, cooling and hot water system; Threshold heating and cooling temperature

2*: Appliances Model

3*: Climate Index based on principal component

4*: Multiple lag output predictions of ambient air temperature

5*: Transmittivity, orientation and insulation thickness

6*: Heating degree hour method

7*: Predict value of temperature, present electricity load, hour and day

8*: Outside temperature and mass flow rate in previous 24 hour, hot water temperature

9*: Highest and Lowest open air temperature

10*: Season, insulation, wall thickness, heat transfer coefficient

11*: Multiple lag of dry bulb temperature and solar radiation

12*: Reference day of each day based on occupancy schedule

13*: Correlated weather data based on reference day and accuracy of calibrated prediction error of previous 2 hour

14*: Occupancy profile represented by space electrical power demand

15*: Clearness of sky, rainfall, cloudiness conditions

16*: Physical and geometrical parameters, hourly cooling load factor

None of these studies has evaluated the transition and time dependent effects of operational power level characteristics of heating plant system and has predicted building heating energy demand in short time horizon (a couple of days). This short term prediction is important to ESCOs for dynamic control of heat plant system. This paper bridges the gap between static and dynamic neural network methods with occupancy profile and operational power level characteristics of heating plant system. It introduces novel pseudo dynamic model, which incorporates time dependent attributes of operational power level characteristics. Their effects on neural network model performances are compared to static neural network for building heating demand. Orthogonal arrays are applied to the proposed pseudo dynamic model for robust design and confirmed the new schedule of occupancy profile and operational heating power level characteristics obtained from ESCOs. The proposed method allows short term horizon prediction (around 4 days with sampling interval of 15 minutes) to make decision (e.g. management of wood power plant) for the ESCOs. The next section describes methodology including scope of study, design of transitional and pseudo dynamic characteristics, neural network model and orthogonal arrays. Finally, a case study is presented and results and discussion are drawn to analyze the performance of different static and pseudo dynamic models along with robustness of proposed pseudo dynamic model for heating demand prediction of the building.

## 2. Methodology

The development and implementation of models proposed in this work are based on collection of real building heating demand, operational heating power level characteristics, climate variables and approximated occupancy profile data (see Appendix A for selection of relevant input variables). An outline of the methodology presented in this paper is shown in figure (1). The input of this methodology is in form of time-series climate and building heating energy data. The other inputs data are occupancy profile and operational heating power level characteristics for working and off-days for 24 hours. Dynamics of building heating demand is also an input to the methodology which includes settling and steady state time and is estimated from real building data. Based on operational heating power level and dynamics of building characteristics, transitional and pseudo dynamic models are designed. Finally, neural networks for static and pseudo dynamic models are designed to predict heating demand in short time horizon (couple of days). For the robustness of pseudo dynamic model, occupancy profile and operational power level characteristics are analyzed for different time intervals to confirm occupancy schedule profile and operation of plant system from the orthogonal arrays. The pseudo dynamic model after optimum orthogonal arrays design is used for final prediction of the building heating demand. Scope of this study, details of transitional and pseudo dynamic model, neural network model and orthogonal arrays are described in section 2.1 - 2.4.

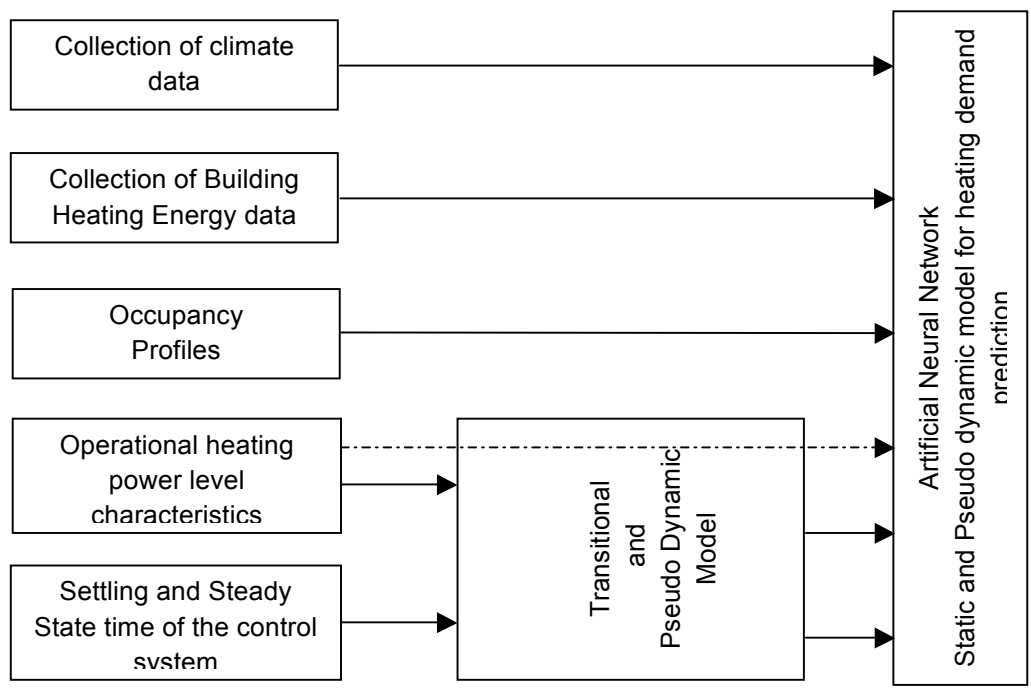

Figure 1: Outline of the proposed methodology on heating demand prediction

## 2.1 Scope of Study

The scope of this paper is heating demand prediction in short time horizon for the large building. The overall objective is to make an energy services decisions (e.g. management of wood power plant) for ESCOs. The assumptions carried for this study are highlighted as:

1. Winter period is studied.
2. Existing building is considered and space heating demand of this building is fed up from a heat network to a central substation. Domestic hot water (DHW) is out of the scope.
3. The heating demand data was recorded in data acquisition system database and thermal comfort inside the building was performed in this database. Thus, the effects of ventilation and air-conditioning on heating are already included in this database.
4. Simple occupancy profile of building is anticipated approximately to assist the ESCOs to schedule their heat production system. In such a system, individual occupant's behavior or precise occupancy profile is not considered. Thus, the modeling constraints are closer to the operational condition of ESCOs to estimate the heat demand.
5. The wind speed and direction are not taken into consideration. This is due to the fact that present weather variables data are taken from data acquisition system but future weather variables values are coming from an atmospheric modeling system which mesh size can be 15 km (as ARPEGE, see [33]), 10 km (as ALADIN, see [34]) or 2.5 km (as AROME, see [35]). In such a case, wind impact on heating demand prediction of a specific building located inside the mesh is very difficult or even impossible to consider for precise effect. Further, heating energy demand is highly dependent on outside temperature and other climate variables have less significant impact on heat energy [36].

## 2.2 Transitional and Pseudo Dynamic Model

The operational heating power level characteristics gives operational features of the plant system, however, they do not give abstract information about transition attributes of operational heating power level which is illustrated through an example in figure (2). The y-axis represents set up power level from the production system and x-axis represents operation schedule.

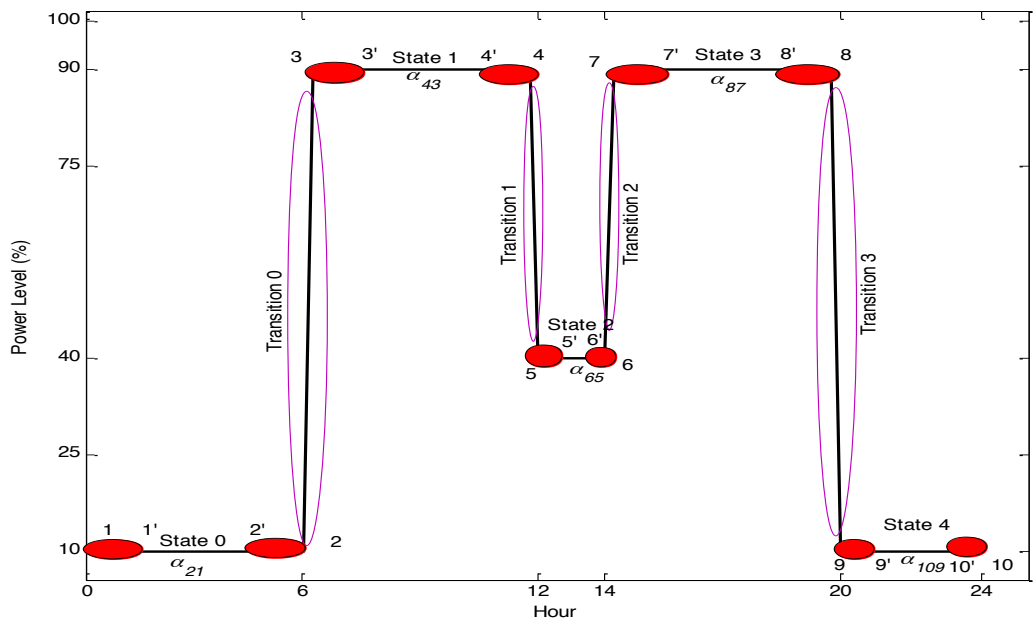

Figure 2: Operational heating power level characteristics of the plant system (for a day)

In figure (2), operational power levels are identified by different states and transition levels and each level has its own significant effects on the operational power level characteristics. State means consistency in the power level from one operation schedule to another and transition means change in power level from one operation schedule to another in heat production system. The transition level 0, 1, 2 and 3 have similar feature of transitional power level characteristics on the overall operational performance, however, power level required for transition from point 2 to 3, point 4 to 5, point 6 to 7 and point 8 to 9 is different for each level. If the power level of state 0, 1, 2, 3 and 4 in operational heating power level characteristics is represented by the $\alpha_{uv}$, then the power required for transition from point $v$ to point $u$ can be represented as $\beta_{uv}$ in the transitional characteristics as shown in figure (3). Thus, the power level transition in transitional characteristics corresponding to operational characteristics can be written as:

$$\begin{aligned} \beta_{uv} &= \beta_{(u-2)(v-2)} + 2\Delta\beta \left| \alpha_{uv} - \alpha_{(u-2)(v-2)} \right|, \forall u = 4,6,8....., v = 3,5,7... \\ \beta_0 & \qquad\qquad\qquad\qquad\qquad , v = 1, u = 2 \end{aligned} \qquad (1)$$

where, $\beta_0$, $\Delta\beta$ and $|\;|$ represents initial power level, step size of transition power level and absolute values respectively. Each level ($\beta_{21}, \beta_{43}, \beta_{65}, \beta_{87}$ and $\beta_{109}$) represents transitional level and depends on the power level of operational characteristics.

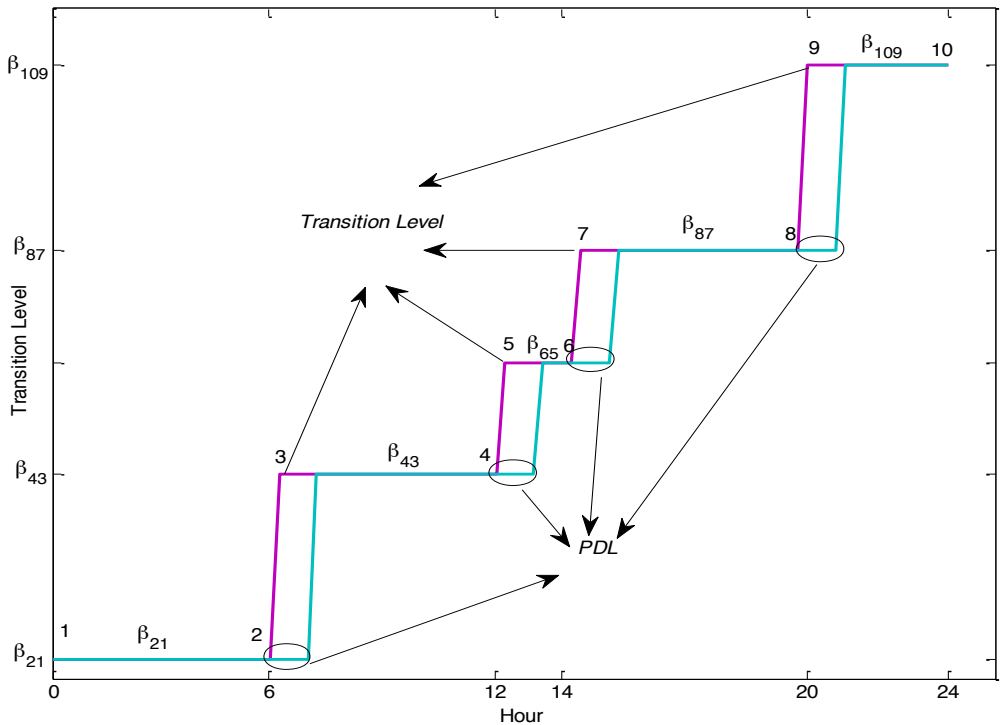

Figure 3: Transitional and Pseudo dynamic characteristics (for a day)

The transitional characteristics explicate the power transition level of operational characteristics, however, dynamic information of power level attributes is still lacking. It means that power content in operational characteristics of figure (2) of point 1-1' is not equal to 2-2'; point 3-3' is not equal to the 4-4'; 5-5' is not equal to 6-6'; 7-7' is not equal to 8-8' and 9-9' is not equal to 10-10'. Dynamic transition information, thus, is necessary in the model which considers dynamic characteristics of the building. The simple first order dynamics of building characteristic is shown in figure (4), where $\tau$ represents time constant.

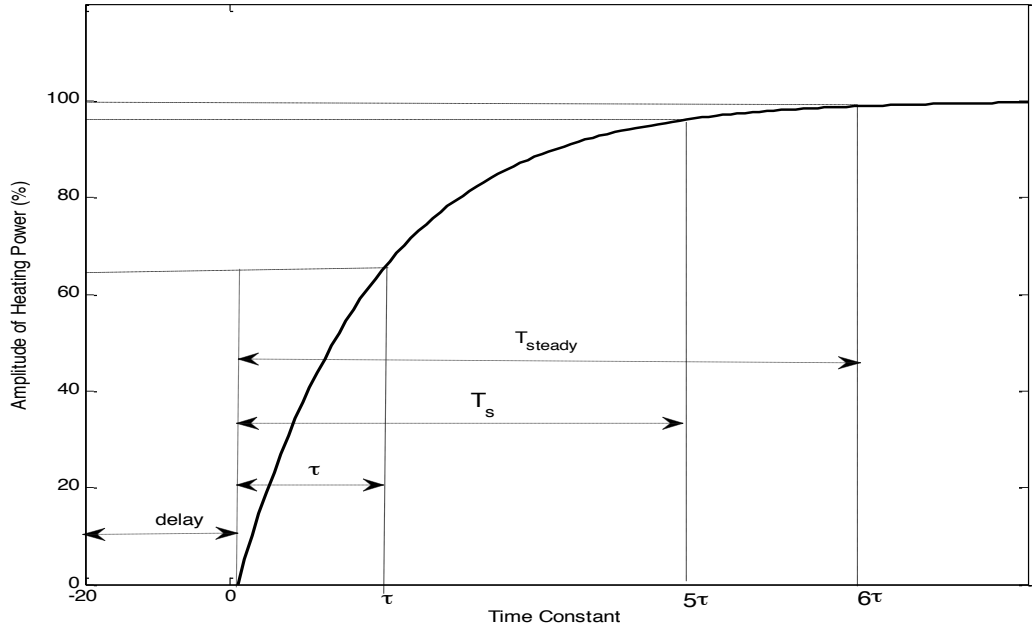

Figure 4: Dynamics of building characteristics

In figure (4), delay represents time it takes from plant system to reach the building for heating operation and after this, power is sufficient to provide heating demand. The $\tau$ represents the 63% of power transferred to the building heating system from plant system. Other dynamics to incorporate is settling time ($T_s$), which is the time elapsed for heating power to reach and remain within the specified error band and equal to [$2\tau$, $5\tau$] and have almost similar behavior like steady state time. The steady state time corresponds to [$3\tau$, $6\tau$]. Thus, $\tau$, settling time ($T_s$) and steady state time ($T_{steady}$) gives information about dynamic characteristics of heating demand. This dynamic information of building, thus, depends on the transitional attributes of power level and this information is not totally dynamic but pertaining to the appearance of dynamic behavior, so pseudo dynamic name is chosen. Thus, pseudo dynamic is just a lag of transitional attribute information and further depends on time constant $\tau$ or range between settling and steady state of the dynamic building heating characteristics. The simplified pseudo dynamic lag (PDL) is calculated from equation (2), where, $ts$ represents the sampling time of building data and $T_u$ represents the new unknown time which lies between settling and steady state time. The concise value of $T_u$ depends on dynamics of the heating demand and pseudo dynamic characteristics can be seen from figure (3), where PDL is pseudo dynamic lag.

$$T_s \leq T_u \leq T_{steady}, \text{where } T_s \in [2\tau, 5\tau]; \quad T_{steady} \in [3\tau, 6\tau]$$

$$\text{PDL} \in \frac{\tau}{ts}[3, 6]$$

(2)

## 2.3 Neural Network Model

The neural network consists of neurons to interconnect the inputs, model parameters and activation function. Each interconnection between the neurons represents model parameters. Input-output mapping in neural network is based on the linear and non-linear activation function. From input and targeted data, model parameters are adjusted to minimize the error i.e. difference between actual values and predicted values produced by the network. Learning/training of data are repeated until there is no significant change in the model parameters and only stops the training. This type of learning approach is called supervised learning since predicted value of the model is guided by actual values.

There are numerous ANN model like Feed-forward Multilayer Perceptron (MLP), Radial Basis Function (RBF) Network, Recurrent Network and Self-Organizing Maps (SOM) [37]. All of these networks have their own learning algorithm to learn and generalize the network. In this paper, MLP is taken as a neural network model since pseudo dynamic model is not fully dynamic (in time behavior). There are two ways of learning mechanism in the neural network: sequential learning and batch learning. In sequential learning, cost function is computed and model parameters are adjusted after each input is applied to the network. In batch learning, all the inputs are fed to the network before model parameters are updated. In batch learning, model parameter adjustment is done at the end of epoch (one complete representation of the learning process) and for this paper, batch learning is carried out.

MLP network consists of three layers: input layer, hidden layer and output layer and there can exist more than one hidden layer. However, according to the Kolmogorov's theorem [38], single hidden layer is sufficient to map the function provided suitable hidden neurons and for this paper, single hidden layer is used as shown in figure (5). The hidden layer assists to solve non-linear separable problems.

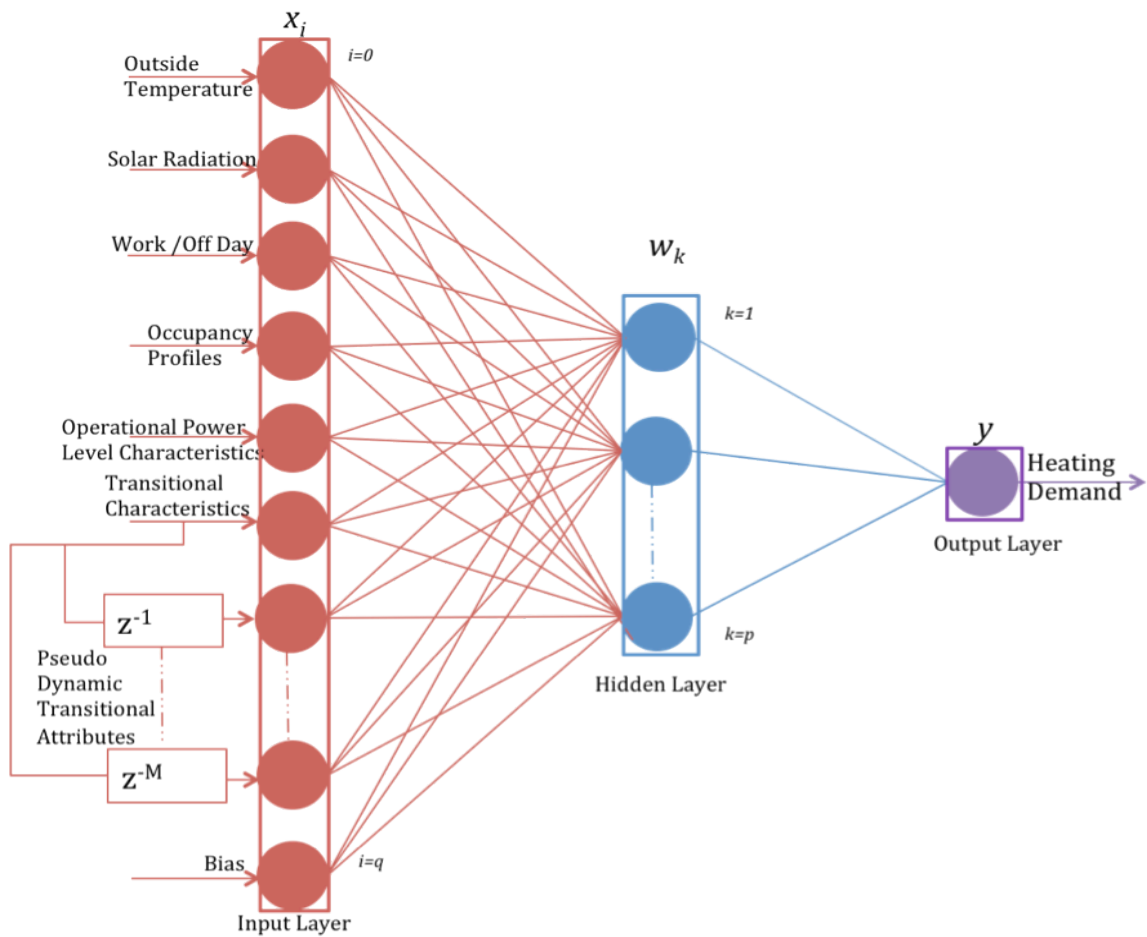

Figure 5: Neural Network Architecture

In figure (5), $x_i$, $w_k$ and $y$ represents input neuron which varies from $i = 0$ to $i = q$, hidden neuron which varies from $k = 0$ to $k = p$ and output neuron respectively. The $z^{-1}$ signifies transition lag of 1 and $z^{-M}$ signifies transition lag corresponding to PDL, where maximum value of M ($M_{max}$) equals to PDL i.e. $M_{max} = \{1, 2, .....PDL\}$. The MLP uses logistic function or hyperbolic tangent as a threshold function in the hidden layer. It has been identified empirically [39] that network using logistic functions tends to converge slower than hyperbolic tangent activation function in the hidden layer during the learning phase. Hyperbolic tangent activation functions is chosen in the hidden layer and pure linear activation function is chosen in the output layer for this paper and hyperbolic tangent function is shown in equation (3), where $\theta^T$ represents model parameter with transpose of matrix. Division of input and output data into learning, validation and testing gives more generalization of model. Learning data sets are used to learn the behavior of input data and to adjust the model parameters. Validation data is used to minimize the overfitting. It is not used to adjust the model parameter but it is used to verify if any increase in accuracy over learning dataset actually yields an increase in accuracy over dataset that has not learned to the network before. Testing data sets are used to confirm the actual prediction from neural network model which is unknown to neural network before. For this paper, data is divided into learning, validation and testing sets. Normalization of input data is also important for faster convergence to achieve desire performance goal. If input data are poorly scaled during learning process, there is a risk of inaccuracy and slower convergence. It is, thus, essential to standardize the input data before applying to neural network. There are various methods for normalization of input and output variable, and for this paper, normalization with zero mean and unit standard deviation is done as shown in equation (4). In equation (4), $\bar{x}$, $X^i$ and $m$ represents mean of input variable, overall vector of input variable and number of datasets respectively and thus, applies similarly for output variable.

$$h(\theta^T, x) = \frac{e^{\theta^T x} - e^{-\theta^T x}}{e^{\theta^T x} + e^{-\theta^T x}} \tag{3}$$

$$X^i = \frac{x^i - \bar{x}}{\sqrt{\frac{1}{m-1} \sum_i (x^i - \bar{x})}} \tag{4}$$

The cost function of MLP network is computed in equation (5):

$$J(\theta) = \frac{1}{2m} \sum_{l=1}^{m} \left[ y^{(l)} - y_a^{(l)} \right]^2 \tag{5}$$

where $y$, $y_a$, $l$ and $J(\theta)$ represents predicted values produced from the network, actual values of given datasets, individual data from $m$ number of datasets and cost function of the neural network model respectively. Further, $y$ of the network is computed as:

$$y = \sum_{k=1}^{p} \theta_k h\left( \sum_{i=0}^{q} \theta_{ki} x_i \right) \tag{6}$$

In order to update the model parameters for a higher degree approximation on unknown non-linear function for learning process, there are different methods as – gradient descent, Newton's method and so on [37]. Gradient descent is too slow for the convergence, and it takes more time to compute the hessian matrix in Newton's method as well. Levenberg-Marquardt algorithm is used for this paper which takes approximation of hessian matrix in the form of Newton's method and model parameter update equation $\theta_{t+1}$ is given as:

$$\theta_{t+1} = \theta_t - \left[L^T L + \mu I\right]^{-1} L^T J(\theta) \qquad (7)$$

In equation (7), hessian matrix is approximated as $[L^T L]$ and gradient is computed as $L^T J(\theta)$, where, $L$ is Jacobian matrix, $J(\theta)$ is vector of cost function, $\theta_t$ is initial model parameter, $\mu$ is suitable chosen scalar and $I$ is identity matrix. Update model parameter, thus, depends on the cost function and scalar value $\mu$.

### 2.3.1 Stopping Criteria

There are different criteria for stopping the neural network model. For this paper, the stopping criteria depend on number of epochs to learn the network, performance goal, maximum range of $\mu$ and maximum failures in the validation. The performance goal (PG) is given as:

$$\text{PG} = 0.01 \sum_{l=1}^{m} y_a^{(l)} \qquad (8)$$

The maximum failures in validation or accuracy over validation datasets is defined to stop the learning process if the accuracy of learning datasets increase and validation accuracy stays same or decrease.

### 2.3.2 Model Performance

Performances of models are characterized by mean square error (MSE) and coefficient of correlation ($R^2$). The MSE and $R^2$ can be calculated as:

$$\text{MSE} = \frac{\sum_{l=1}^{m} \left[y^{(l)} - y_a^{(l)}\right]^2}{m} \qquad (9)$$

$$R^2 = \frac{\sum_{l=1}^{m} \left[y^{(l)} - y_a^{(l)}\right]^2}{\sum_{l=1}^{m} \left(y_a^{(l)}\right)^2} \qquad (10)$$

### 2.3.3 Degree of Freedom Adjustment

One of the issues of neural network model is over learning of the network. With increase of hidden neurons, model performance can be increased, but, it will lead neural network to over learning. Validation accuracy and degree of freedom (DOF) adjustments are done in this paper to avoid over fitting. Number of learning equations that model could deliver are given by equation (11), where $L_e$ is learning equations of the network and $L_y$ is length of vector output neurons ($y$), and in this case equal to 1 since there is only heating demand load.

$$L_e = m * L_y \qquad (11)$$

The number of model parameters for a single hidden layer MLP neural network are given by the equation (12), where $L_\theta$, $L_x$ and $L_w$ represents number of model parameters, vector length of input neurons ($x_i$) and vector length of hidden neurons ($w_k$) respectively.

$$L_\theta = (L_x + 1) * L_w + (L_w + 1) * L_y \tag{12}$$

DOF of neural network model is the difference between number of learning equations and number of model parameters in the network. It should be always >>1 and depends on the optimum size of hidden neurons. DOF and maximum hidden neurons are given by equation (13) and (14), where, $\delta$ represents the scalar constant value and depends on DOF required for design and $W_{max}$ is the maximum hidden neurons.

$$\text{DOF} = L_e - L_\theta \tag{13}$$

$$W_{max} \cong \frac{1}{\delta} \frac{(L_\theta - L_y)}{(L_x + L_y + 1)} \tag{14}$$

Modified performance goal according to degree of freedom adjustment is given as:

$$\text{PG} = \frac{0.01 \, \text{DOF} \sum_{l=1}^{m} y_a^{(l)}}{L_e} \tag{15}$$

Model performance is also further modified based on degree of freedom adjustment. The modified MSE and R² can be calculated as:

$$\text{MSE}_{modified} = \frac{L_e \sum_{l=1}^{m} \left[ y^{(l)} - y_a^{(l)} \right]^2}{\text{DOF} * m} \tag{16}$$

$$R^2_{modified} = \frac{L_e \sum_{l=1}^{m} \left[ y^{(l)} - y_a^{(l)} \right]^2}{\text{DOF} \sum_{l=1}^{m} \left( y_a^{(l)} \right)^2} \tag{17}$$

For each hidden neurons, optimal $\text{MSE}_{modified}$ and maximum $R^2_{modified}$ for learning and validation are calculated from the different initialized random parameters. For different number of hidden neurons, $R^2_{modified}$ and $\text{MSE}_{modified}$ for each model is performed for learning and validation, and based on it, optimal configuration of model is identified for the final prediction.

### 2.4 Orthogonal Arrays

It is essential to know whether schedule of occupancy profile and operational characteristics obtained from ESCOs is reliable for the robust design of pseudo dynamic model. Occupancy profile and operational characteristics transition period, thus, plays an important role in the model performance and if all these transition period are consider for finding the best robust model, it takes long time to compute. Orthogonal arrays (OA) identify the main effects with minimum number of trials to find the best design. These are applied in various fields: mechanical and aerospace engineering [40], electromagnetic propagation [41] and signal processing [42] for the robust design model.

The orthogonal array allows the effect of several parameters to find best design with given different levels of parameters. It can be defined as matrix with column representing number of parameters with different settings to be studied and rows representing number of experiments. In orthogonal arrays, parameters are called factors and parameter settings are called levels. In general, $OA(N,k,s,t)$ is used to represent the orthogonal arrays, where $N$, $k$, $s$ and $t$ represents number of experiments, number of design parameters, number of levels and strength. There are different methods as Latin square [43]; Juxtaposition [44]; Finite geometries [45] etc... to create orthogonal arrays with different strength and levels. Orthogonal arrays with different number of design parameter, level, and strength are available from OA databases or libraries. The orthogonal arrays used for this paper is taken from OA library [46].

## 3. Case Study

The methodology is applied for case study at Ecole des Mines de Nantes, French Institution. The building has floor area of 25,000 m². It has 600 students and 200 employees. The building consists of 120 research and administration rooms, 30 class rooms, 3 laboratories, and 8 seminar halls. Class rooms have different sizes and can accommodate to 18 to 28 students. The 2 big seminar halls can be occupied by 250 students and 6 small seminar halls can be occupied by 80 students. Each floor area of the laboratory is 600 m².

The data is taken from data acquisition system and consists of day/month/time, solar radiation, outside air temperature and heating demand from mid of January to February 2013 with sampling interval of 15 minutes. The 70% of data (outside temperature, solar radiation and heating demand as shown in figure 5) are used for learning phase i.e. $m$ in mathematical equation in neural network, see section 2.3, equivalent to 19 days with 15 minute sampling time, and each 15% of data (4 days with 15 minute sampling time) is used for validation and testing phase. Outside temperature taken for this study has minimum, average and maximum value of 1.2 $^0$C, 8.95 $^0$C and 15.3 $^0$C respectively. Global solar radiation has an average and maximum value of 7 W/m² and 438 W/m² respectively.

The simplified/theoretical occupancy profile and operational heating power level characteristics for working and off-days for 24 hours is shown in figure (6) and (7).

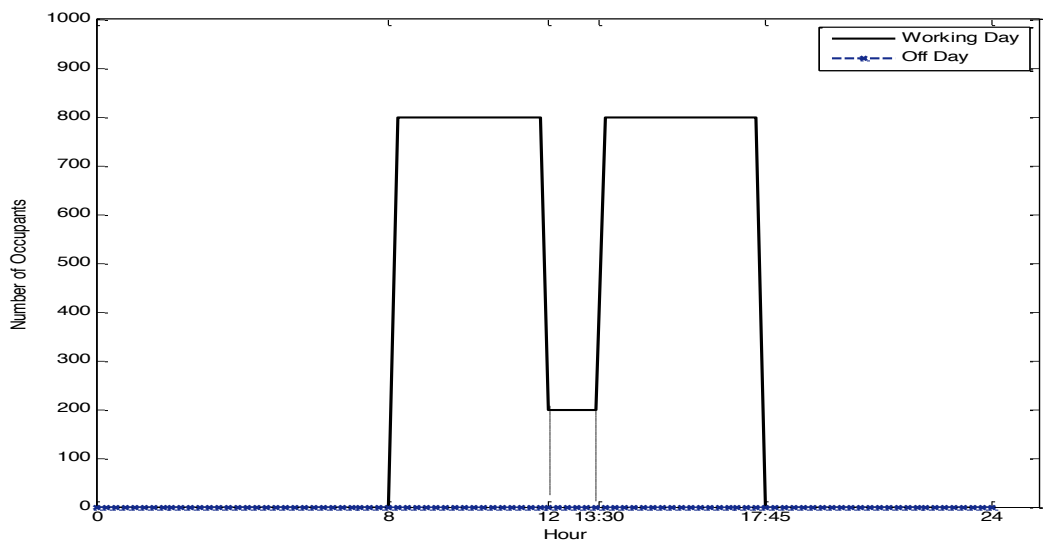

Figure 6: Occupancy profiles for working and off-day

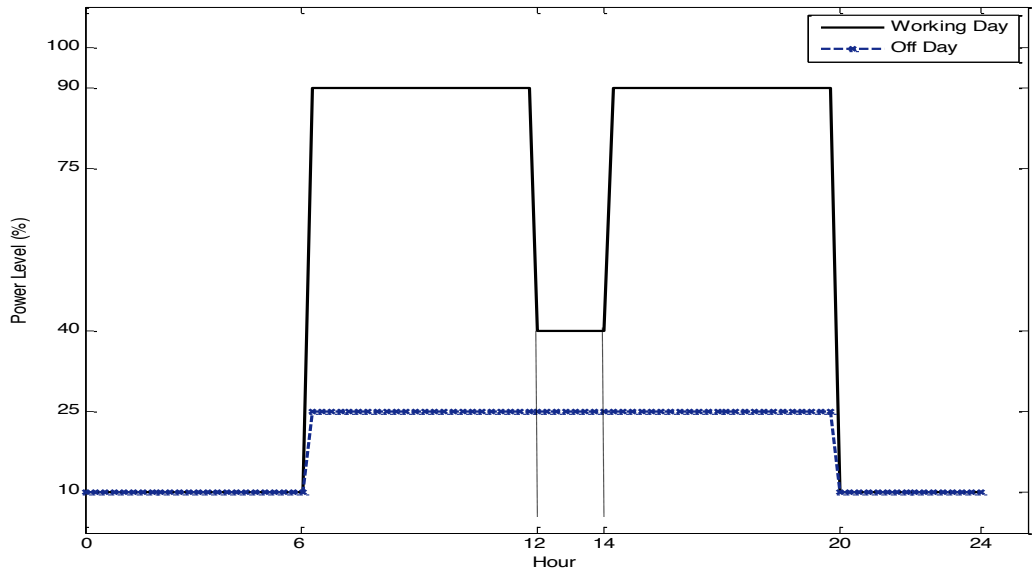

Figure 7: Operational heating power level characteristics for working and off-day

Power demand and occupancy profile during working day is depicted from figure (8). From figure (8), occupancy profile almost gives information about power demand characteristics, however, from 18 hour onwards, power demand characteristics is not accordance with occupancy profile. Thus, it further shows that simplified occupancy profile is not enough to characterize the heating demand.

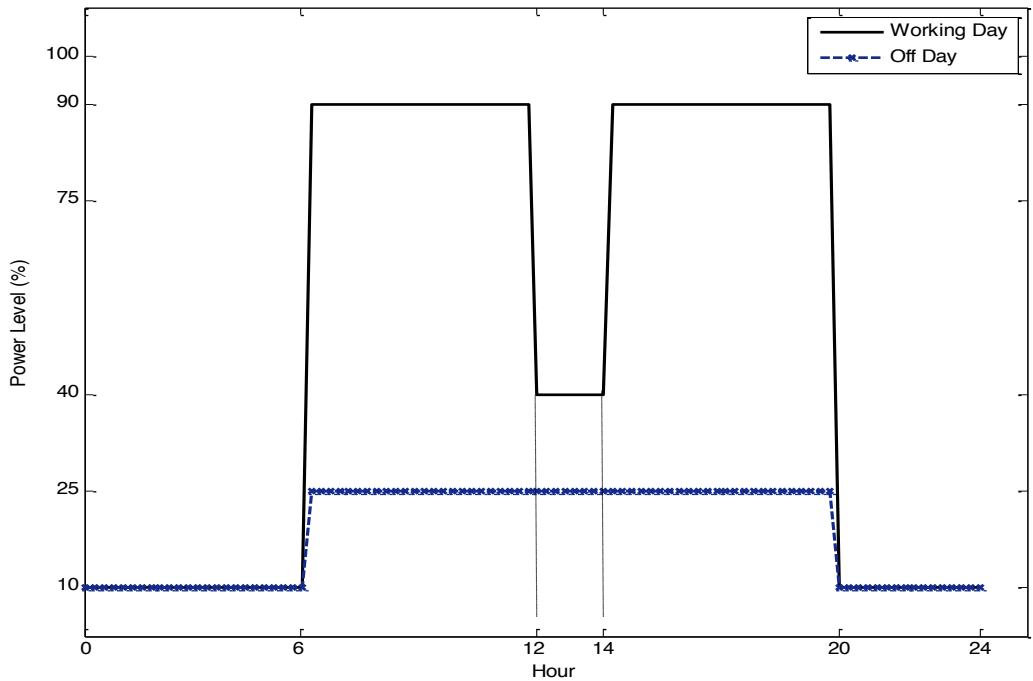

Figure 8: Heating power demand and occupancy profile during working days

Different neural network models are designed based on climate variables (outside temperature and solar radiation), work/off day information, occupancy profile and operational characteristics as shown in figure (5). For this case study, 10 represent working day and 5 represent off day information (work/off day) to the input of neural network model. Static neural network model 1 consists of operational characteristics and occupancy profile, external temperature and solar radiation as input variables and heating power demand as an output variable, and thus, vector length of input neurons ($L_x$) in equation (12) equals to 5. Model 2 comprises additional transitional characteristics in model 1 and vector length of input neurons ($L_x$) in equation (12) equal to 6. For this case study the sampling time ($ts$) of real building data is 15 minutes, settling time ($T_s$) is estimated approximately 45 minutes and steady state time ($T_{steady}$) is approximately 1 hour. The PDL, thus, is calculated from equation (2), where PDL corresponds to settling and steady state time is nearly equal to 3 and 4 respectively. Since pseudo dynamic model depends on transition lag of operational heating power level and building dynamic characteristics, PDL is varied from 3-4, and to understand the phenomena of pseudo dynamic lag, PDL is varied from 1-4. Model 3 comprises model 2 with additional parameters of one PDL i.e. i.e. $L_x$ equals to 7; model 4 consists model 2 with additional parameters of two PDL i.e. $L_x$ equals to 8; model 5 includes model 2 with additional parameters of three PDL i.e. $L_x$ equals to 9 and model 6 comprises model 2 with additional parameters of four PDL in the transitional characteristics i.e. $L_x$ equals to 10. Transitional and pseudo dynamic characteristic with four lags during working day is shown in figure (9). Transition level in figure (9) is calculated from equation (1) and for this case study, 25 is chosen for each $\beta_0$ and $\Delta\beta$. In figure (9), lag 0 means static model which contains transition attributes, lag 1 means pseudo dynamic model with transition lag 1 (PDL=1), lag 2 means pseudo dynamic model with transition lag 2 (PDL=2) and so on. Further, effects of transitional and pseudo dynamic effects on the heating demand can be understood from figure (10). It is clear that the information hidden in heating demand which climate variables could not answer can be justify from transitional and pseudo dynamic attributes of operational characteristics. The summary of models is shown in table (2).

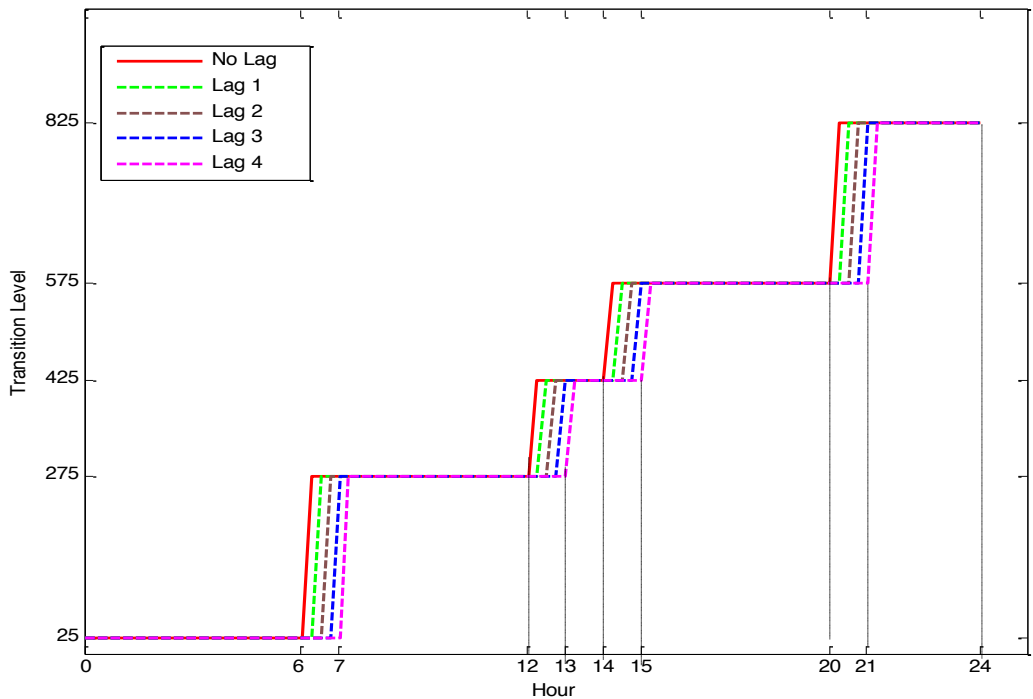

Figure 9: Transitional and pseudo dynamic characteristics during working day

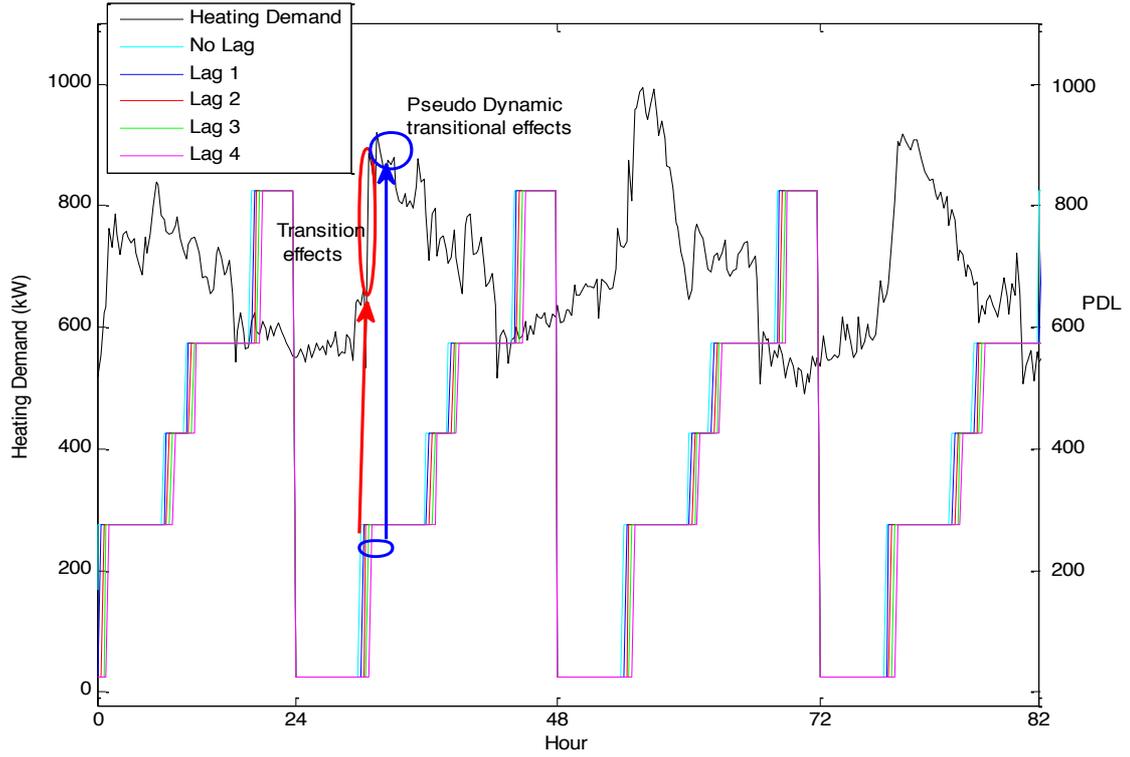

Figure 10: Pseudo dynamic transitional effects on heating demand

Table 2: Summary of models

| Model No. | Type of Model | Input Variables | Remarks |
|---|---|---|---|
| Model 1 | Static | Climates, occupancy profile and operational characteristics | No Lag |
| Model 2 | Static | Model 1 with transitional characteristics | No Lag |
| Model 3 | Pseudo Dynamic | Model 2 with pseudo dynamic transition in dead band | Lag 1 |
| Model 4 | Pseudo Dynamic | Model 2 with pseudo dynamic transition in t | Lag 2 |
| Model 5 | Pseudo Dynamic | Model 2 with pseudo dynamic transition in settling time | Lag 3 |
| Model 6 | Pseudo Dynamic | Model 2 with pseudo dynamic transition in steady state time | Lag 4 |

For each model, cost function $J(\theta)$ in equation (5) is computed iteratively up to 1000 for each of the minimum and maximum number of hidden neurons. The maximum number of hidden neurons is calculated from equation (14), where $\delta$ is chosen 8 as it gives the flexibility in the degree of model parameters. Thus, three minimum hidden neurons are chosen as 3 for this case study. Hidden neurons length ($L_w$), thus, is varied from 3 to $W_{max}$. Performance of model at each iteration (number of epochs) is computed from equation (16) and (17) and model parameters are updated based on equation (7), where initial value of $\mu$ is chosen as 0.01 and its value is increased with a factor of 10 and decreased with a factor of 0.1. The maximum value of $\mu$ is chosen as 1e10. Neural network model in this study will be stopped if the number of epochs reached to 1000 and performance goal reached the value given by equation (15).

Under the scope of study (see subsection 2.1), the accuracy on the number of occupants are not relevant, however, it is essential to know inside the sampling time, when the staff and students come and leaves the buildings. It is necessary to check occupancy and operational power level characteristics provided by ESCOs are right or not for robust design model. And, the main controlling factors for robust design model are the transition schedule of occupancy and operational characteristics. From figure (6), it is clear that there is no transition of occupancy during off-day, but there is transition of occupancy during the interval at 8 hour, 12 hour, 13:30 hour and 17:45 hour and these are represented by t1, t2, t3 and t4 factors respectively. Similarly, there is a transition of operational characteristics for working and off day as shown in figure (7) and these transition factors are represented by t5, t6, t7 and t8 for working day for 6 hour, 12 hour, 14 hour and 20 hour; t9 and t10 for off day for 6 hour and 20 hour. Since the sampling interval taken for this case study is 15 minutes, three levels are used for orthogonal arrays so that the model will represent the 15 minutes ahead and before from occupancy and operational characteristics schedule period. The summary of control factors and their levels are shown in table (3), where OSW represents occupancy schedule at work day, OCSW represents operational characteristics schedule at work day and OCSO represent operational characteristics schedule at off day.

Table 3: Summary of control factors and their levels

| Factors | Levels | | |
|---|---|---|---|
| | 1 | 2 | 3 |
| OSW at 8 hour (f1) | t1-15 min | t1 | t1+15 min |
| OSW at 12 hour (f2) | t2-15 min | t2 | t2+15 min |
| OSW at 13:30 hour (f3) | t3-15 min | t3 | t3+15 min |
| OSW at 17:45 hour (f4) | t4-15 min | t4 | t4+15 min |
| OCSW at 6 hour (f5) | t5-15 min | t5 | t5+15 min |
| OCSW at 12 hour (f6) | t6-15 min | t6 | t6+15 min |
| OCSW at 14 hour (f7) | t7-15 min | t7 | t7+15 min |
| OCSW at 20 hour (f8) | t8-15 min | t8 | t8+15 min |
| OCSO at 6 hour (f9) | t9-15 min | t9 | t9+15 min |
| OCSO at 20 hour (f10) | t10-15 min | t10 | t10+15 min |

Thus, there are 10 factors and 3 levels that govern the robustness of the model and if the full factorials are used to generalize the model, it takes $3^{10}$ = 59049 experiments. The orthogonal arrays reduce the number of experiments to 729 with 5 strengths. OA (729,10,3,5) is applied to the proposed pseudo dynamic model in this case study.

**4. Result and Discussion**

Optimal configuration of the model is based on maximum $R^2_{modified}$ and minimum $MSE_{modified}$ from different random initialized parameters. For each hidden neurons in the model, five random initialized parameters is assigned for learning phase and based on it, the neurons with minimum $MSE_{modified}$ and maximum $R^2_{modified}$ for learning and validation are chosen from random initialized parameters. Optimal configuration of each model is chosen from maximum $R^2_{modified}$ and minimum $MSE_{modified}$ model performance from learning and validation datasets for different hidden neurons. Figure (11) and (12) shows $R^2_{modified}$ and $MSE_{modified}$ performance for learning, validation and testing for different hidden neurons sizes of model 5 and from this optimal configuration is chosen from the best performance model. It is clear from figure (11) and (12) that the maximum $R^2_{\text{mod}ified}$ and minimum $\text{MSE}_{\text{mod}ified}$ performance is achieved in hidden neuron size 13 and which is the optimal configuration of the model. It can also be noticed that although $R^2$ testing performance increases for hidden neuron size 15, $R^2$ for validation and learning does not increase optimally. The model 5 is just an example and similarly, the process is repeated for each model to find the optimal configuration of the neural network model. The optimal configurations of the different neural network model are summarized in table (4).

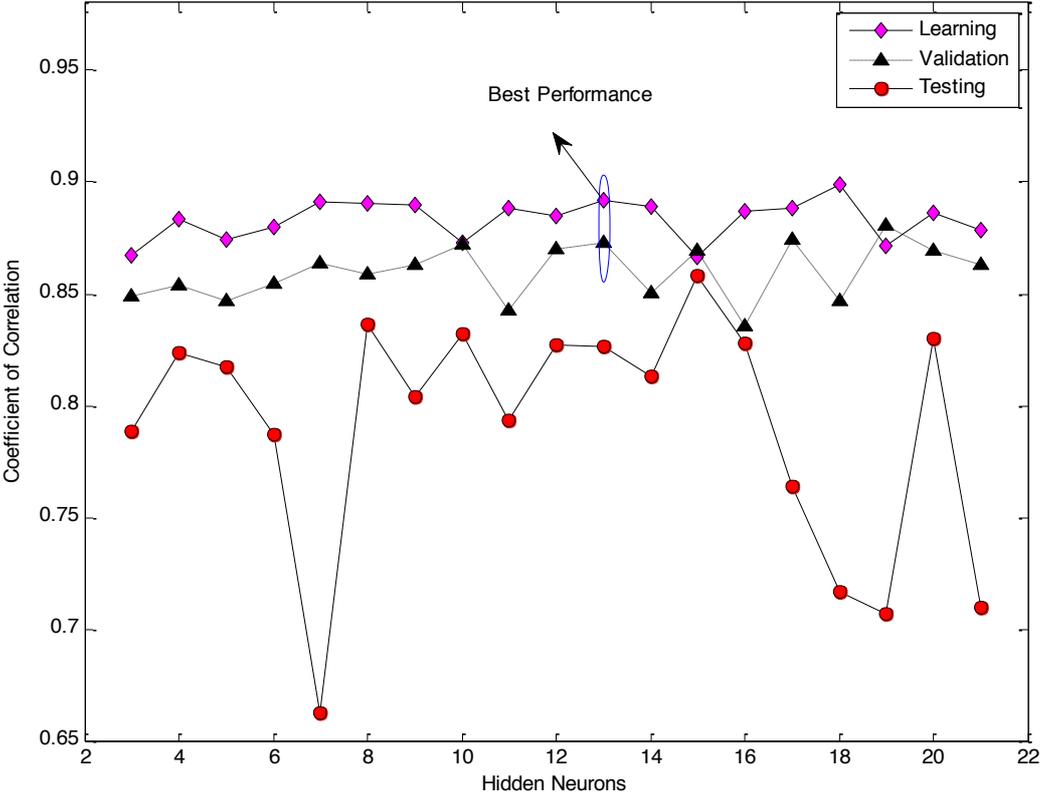

Figure 11: Coefficient of correlation performance (Model 5)

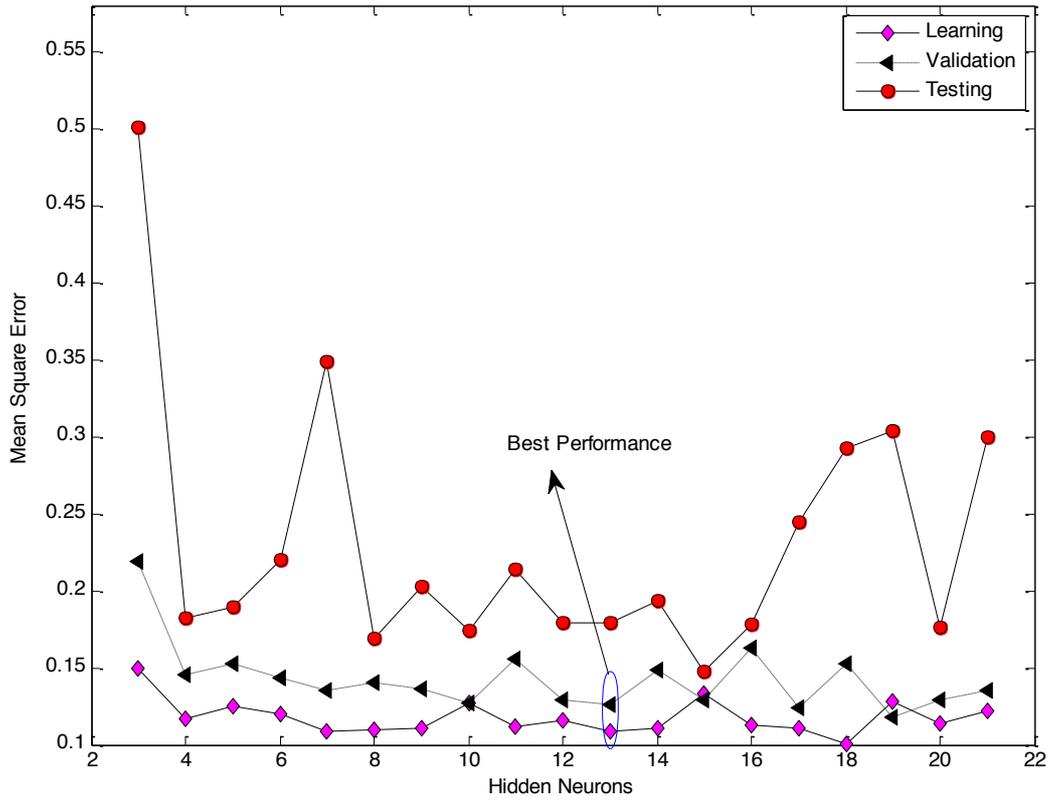

Figure 12: Mean Square Error performance (Model 5)

Table 4: Optimal configuration of models

| Model | Hidden Neurons | Coefficient of Correlation | | | Mean Square Error | | |
|---|---|---|---|---|---|---|---|
| | | Learning | Validation | Testing | Learning | Validation | Testing |
| Model 1 | 10 | 0.82 | 0.81 | 0.61 | 0.18 | 0.18 | 0.40 |
| Model 2 | 19 | 0.87 | 0.85 | 0.80 | 0.13 | 0.15 | 0.21 |
| Model 3 | 7 | 0.88 | 0.86 | 0.75 | 0.12 | 0.14 | 0.25 |
| Model 4 | 9 | 0.89 | 0.87 | 0.82 | 0.12 | 0.13 | 0.18 |
| Model 5 | 13 | 0.89 | 0.87 | 0.83 | 0.11 | 0.13 | 0.18 |
| Model 6 | 9 | 0.89 | 0.87 | 0.85 | 0.11 | 0.13 | 0.15 |

Table (4) shows that with static neural network model 1, best $R^2_{modified}$ for learning and validation can be obtained up to 0.82 and 0.81. From this, it is clear that occupancy profile and operational characteristics are not enough to determine and generalize the unknown function of the building heating demand. As transitional attributes of operational characteristic is introduced in model 2, $R^2_{modified}$ model performance increases significantly from 0.82 to 0.87 for learning phase and from 0.81 to 0.85 for validation phase and correspondingly $MSE_{modified}$ decreases in contrast to model 1. Pseudo dynamic transitional attributes in model 3 and time constant $\tau$ in model 4 leads increase in model performance. Further, dynamics of settling time and steady state plays an important role in characterizing the neural network model. It is seen that $R^2_{modified}$ performance increases from 0.87 to

0.89 for learning and 0.85 to 0.87 for validation in model 5 compare to model 2 although transition attributes is introduce in model 2. In addition, hidden neuron size is also reduces from 19 to 13. Moreover, it is distinguish that learning and validation performances remained the same in the model 6 compared to model 5. The optimal choice of the model, thus, lies in between settling and steady state time.

It can be further view that model 5 and model 6 show reasonable and consistent model performances. However, minimum hidden neuron size and maximum learning criteria is essential for the overall network generalization. Since the hidden neurons size decreases from 13 to 9 and model performance $R^2_{modified}$ remained the same (0.89) in model 6 comparing to model 5, model 6 is chosen as the best configuration of the overall models. The optimal choice of the model 5 and model 6 can be delineated by the error in percentage of energy consumption (kWh) in actual and prediction for the learning and validation phase. Heating energy consumption error in actual and prediction in learning phase in Model 6 is 0.02% compare to 0.32% in Model 5. For validation phase, heating energy consumption error is 2.39% in Model 6 compare to 2.57% in Model 5. From this energy consumption error, it is clear that there is a small heating energy consumption error in Model 6 compare to Model 5 during the learning and validation phase. So, one can conclude that Model 6 can be chosen as optimal configuration of the overall model. The model 6, thus, bridges the gap between static and dynamic neural network model in the sense that it is better than static model and increases the performance comparable to dynamic neural network model.

For the robustness of pseudo dynamic model, orthogonal arrays are applied to determine the highest coefficient of correlation for learning and validation for the optimum 9 hidden neuron size of model 6. Table (5) shows OA(729,10,3,5) and coefficient of correlation for learning and validation phase. It is clear from table (5) that the schedule taken from the ESCOs is from experiment 1 and from the orthogonal arrays, the optimal schedule that fits the best for model 6 is experiment 398. The orthogonal arrays, thus, ensures that there is transition in occupancy in 7:45 hour, 12 hour, 13:45 hour and 18 hour instead of 8 hour, 12 hour, 13:30 hour and 17:45 hour period in the existing case respectively. There is also a transition in 5:45 hour, 11:45 hour, 14 hour and 17:45 hour instead of 6 hour, 12 hour, 14 hour and 17:45 hour for working day; 5:45 hour and 20 hour instead of 6 hour and 20 hour in off days for operational characteristics. The coefficient of correlation after the orthogonal array design is 0.90 for learning, 0.88 for validation and 0.86 for training phase. Nevertheless, other issue of overall model is that it is difficult to increase the coefficient of correlation beyond 0.90 and this is due to the sampling time of 15 minutes. With short sampling time, it is very difficult to learn the datasets which changes in 15 minutes sample, nonetheless, for good generalization of the model, $R^2_{modified}$ value of 0.90 during the learning phase is always acceptable.

Coefficient of correlation of linear regression obtained from neural network model in the actual and prediction of heating demand for learning, validation and testing phase of Model 6 after optimum orthogonal array design are 0.95, 0.95 and 0.93 respectively. The prediction of heating demand for model 6 after optimum orthogonal array design during validation phase is shown in figure (13). Prediction gives the power heating demand and the area under the curve gives the heating energy demand. From figure (13), it is clear that heating demand tremendously increases approximately 990 kW during third and fourth day and pseudo dynamic model is able to predict and learn the behavior. However, there is a fluctuation in the power demand in the morning for each consecutive 4 days and it is difficult to learn datasets which transits rapidly in actual power demand. The prediction of heating demand for model 6 during testing phase after optimum orthogonal array design is shown in figure (14). It is vivid that pseudo dynamic model is able to predict heating demand, however during the third day, the pseudo dynamic model is not able to meet 1.1 MW of heating demand. This is due to the fact that neural network does not learn this threshold maximum heating demand in the learning phase as this kind of information is not available in the database. This data, thus, needs to be improved in the learning phase through feature extraction techniques. Nonetheless, pseudo dynamic model (model 6) prediction is in accordance to the actual target except for some rapid transits in the actual target. To sum up, pseudo dynamic transition attributes in model 6 after orthogonal array design leads best prediction of heating demand.

Table 5: OA(729,10,3,5) and coefficient of correlation for learning and validation for model 6

| Experiment | Element | f1 | f2 | f3 | f4 | f5 | f6 | f7 | f8 | f9 | f10 | Coefficient of Correlation | | |
|---|---|---|---|---|---|---|---|---|---|---|---|---|---|---|
| | | | | | | | | | | | | Learning | Validation | Testing |
| 1 | | 2 | 2 | 2 | 2 | 2 | 2 | 2 | 2 | 2 | 2 | 0.89 | 0.87 | 0.85 |
| 2 | | 1 | 2 | 2 | 2 | 2 | 2 | 2 | 1 | 1 | 1 | 0.89 | 0.88 | 0.81 |
| 3 | | 3 | 2 | 2 | 2 | 2 | 2 | 2 | 3 | 3 | 3 | 0.90 | 0.86 | 0.76 |
| 4 | | 2 | 1 | 2 | 2 | 2 | 2 | 3 | 2 | 1 | 3 | 0.89 | 0.86 | 0.79 |
| 5 | | 1 | 1 | 2 | 2 | 2 | 2 | 3 | 1 | 3 | 2 | 0.89 | 0.87 | 0.78 |
| 6 | | 3 | 1 | 2 | 2 | 2 | 2 | 3 | 3 | 2 | 1 | 0.89 | 0.88 | 0.79 |
| 7 | | 2 | 3 | 2 | 2 | 2 | 2 | 1 | 2 | 3 | 1 | 0.89 | 0.87 | 0.83 |
| 8 | | 1 | 3 | 2 | 2 | 2 | 2 | 1 | 1 | 2 | 3 | 0.89 | 0.87 | 0.84 |
| 9 | | 3 | 3 | 2 | 2 | 2 | 2 | 1 | 3 | 1 | 2 | 0.90 | 0.86 | 0.85 |
| 10 | | 2 | 2 | 1 | 2 | 2 | 2 | 3 | 1 | 2 | 1 | 0.89 | 0.87 | 0.76 |
| 11 | | 1 | 2 | 1 | 2 | 2 | 2 | 3 | 3 | 1 | 3 | 0.89 | 0.87 | 0.67 |
| 12 | | 3 | 2 | 1 | 2 | 2 | 2 | 3 | 2 | 3 | 2 | 0.89 | 0.87 | 0.67 |
| …. | | . | . | . | . | . | . | . | . | . | . | . | . | . |
| … | | . | . | . | . | . | . | . | . | . | . | . | . | . |
| 394 | | 2 | 3 | 1 | 3 | 1 | 1 | 3 | 3 | 3 | 3 | 0.89 | 0.87 | 0.80 |
| 395 | | 1 | 3 | 1 | 3 | 1 | 1 | 3 | 2 | 2 | 2 | 0.90 | 0.87 | 0.76 |
| 396 | | 3 | 3 | 1 | 3 | 1 | 1 | 3 | 1 | 1 | 1 | 0.90 | 0.87 | 0.76 |
| 397 | | 2 | 2 | 3 | 3 | 1 | 1 | 2 | 2 | 2 | 3 | 0.89 | 0.88 | 0.81 |
| 398 | | 1 | 2 | 3 | 3 | 1 | 1 | 2 | 1 | 1 | 2 | 0.90 | 0.88 | 0.86 |
| 399 | | 3 | 2 | 3 | 3 | 1 | 1 | 2 | 3 | 3 | 1 | 0.90 | 0.87 | 0.70 |
| 400 | | 2 | 1 | 3 | 3 | 1 | 1 | 3 | 2 | 1 | 1 | 0.90 | 0.87 | 0.77 |
| 401 | | 1 | 1 | 3 | 3 | 1 | 1 | 3 | 1 | 3 | 3 | 0.89 | 0.88 | 0.84 |
| 402 | | 3 | 1 | 3 | 3 | 1 | 1 | 3 | 3 | 2 | 2 | 0.87 | 0.88 | 0.74 |
| …. | | . | . | . | . | . | . | . | . | . | . | . | . | . |
| … | | . | . | . | . | . | . | . | . | . | . | . | . | . |
| 725 | | 1 | 1 | 3 | 3 | 3 | 3 | 2 | 1 | 2 | 3 | 0.89 | 0.87 | 0.80 |
| 726 | | 3 | 1 | 3 | 3 | 3 | 3 | 2 | 3 | 1 | 2 | 0.90 | 0.87 | 0.84 |
| 727 | | 2 | 3 | 3 | 3 | 3 | 3 | 3 | 2 | 2 | 2 | 0.90 | 0.87 | 0.80 |
| 728 | | 1 | 3 | 3 | 3 | 3 | 3 | 3 | 1 | 1 | 1 | 0.89 | 0.88 | 0.78 |
| 729 | | 3 | 3 | 3 | 3 | 3 | 3 | 3 | 3 | 3 | 3 | 0.89 | 0.88 | 0.61 |

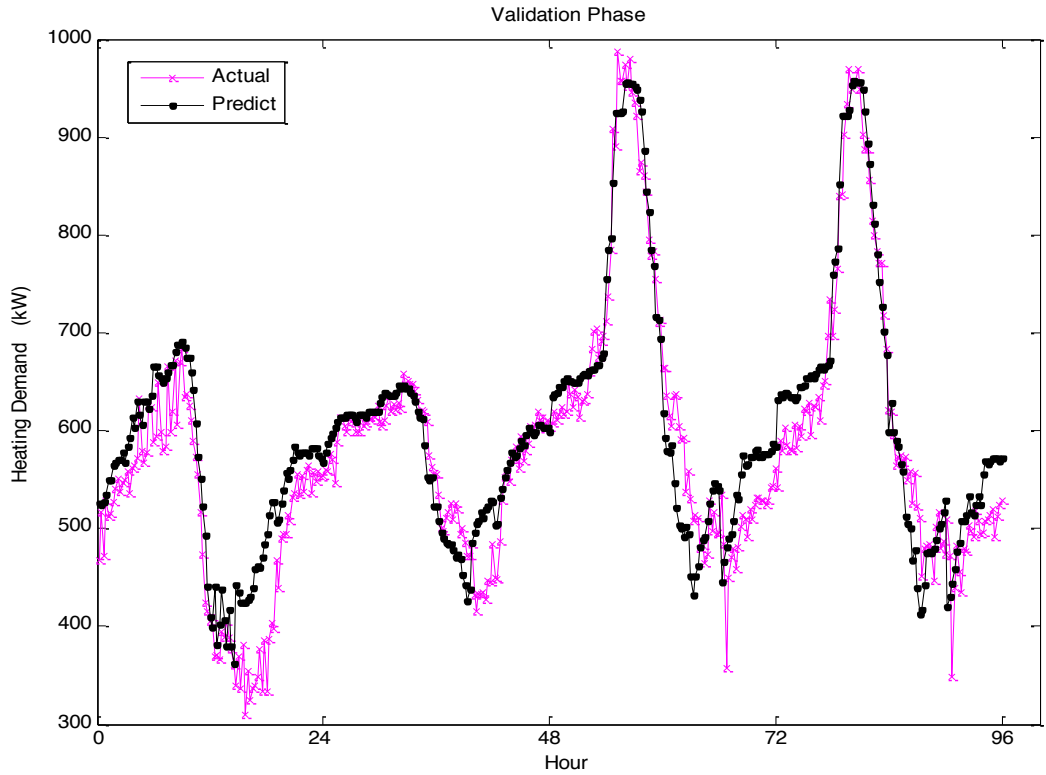

Figure 13: Prediction of heating demand in model 6 during validation phase (after optimum orthogonal array design)

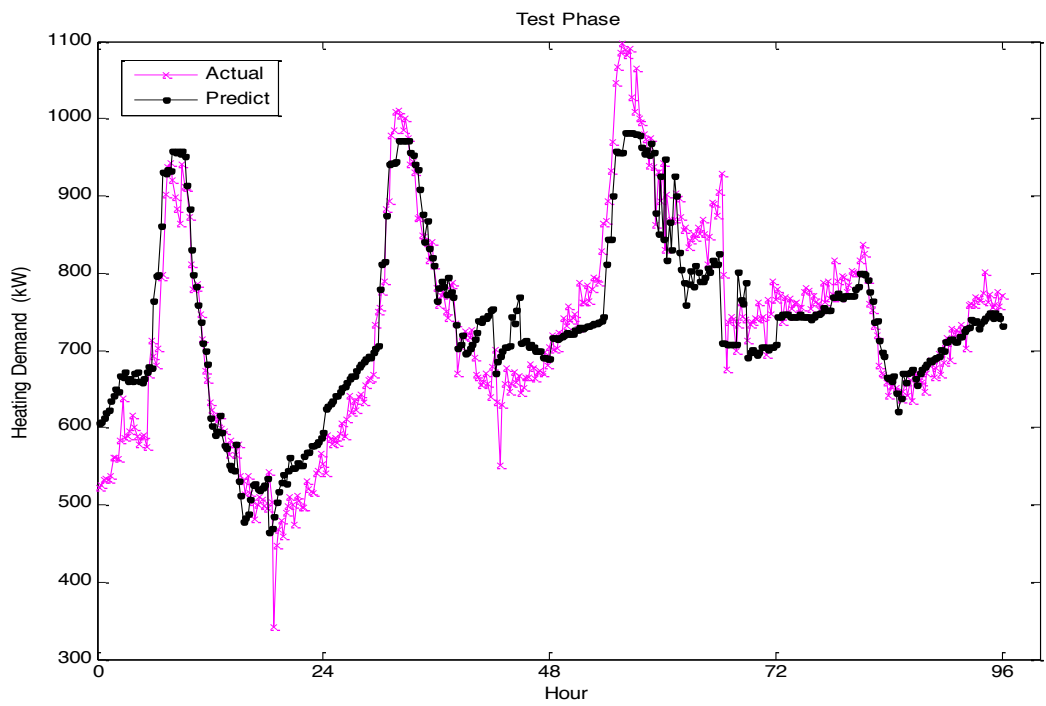

Figure 14: Prediction of heating demand in model 6 during testing phase (after optimum orthogonal array design)

## 4. Conclusion

This paper introduces pseudo dynamic transitional model for the building heating demand prediction in a short time horizon using artificial neural network. Occupancy profile and operational heating power level characteristics are included in the model. Dynamic characteristic of the building is included in the model for the determination of pseudo dynamic transition lag. Settling time and steady state time of the heating demand give an increment in precision of the model, however, choice of model depends on their actual time between settling and steady state. The results were based on case study where occupancy profile is already known and results may vary for more fluctuating occupancy buildings. Coefficient of correlation increases from 0.82 to 0.89 for learning, 0.81 to 0.87 for validation and 0.61 to 0.85 for testing in pseudo dynamic comparing to static neural network model. Also, the size of hidden neuron is further reduced, which reduces complexities and increases generalization of the model. Moreover, minimum energy consumption error is achieved in pseudo dynamic model as 0.02% for learning and 2.57% for validation phase. Further, orthogonal array is applied to optimal pseudo dynamic model to confirm the schedule of occupancy profile and operational level characteristics, and robustness of the model. The orthogonal array design leads to the increases in coefficient of correlation in pseudo dynamic model and confirmed the new schedule of the occupancy profile and operational level characteristics. The major contribution of this paper, thus, is the introduction of transition and novel time dependent attributes of operational heating power level characteristics, which is the dominant factor for building heating demand. Also, orthogonal array design in the model makes flexibility in cross checking the schedule of occupancy profile and operational heating power level characteristics obtained from ESCOs to design the robust model. The prediction is in short time horizon (4 days) with sampling interval of 15 minutes and thus useful for dynamic control of building heating demand.

Further, research will be focused towards the feature extraction of data before learning phase of the neural network so that abnormalities in the data can be corrected in the learning phase. Also adaptive and real time learning criteria with seasonal behaviour will be studied.

**Acknowledgement**

This research has been done in collaboration with Ecole des Mines, Nantes, Technische Universiteit Eindhoven and VEOLIA Environnement Recherche et Innovation, funded through Erasmus Mundus Joint Doctoral Programme SELECT+, the support of which is gratefully acknowledged.

Appendix A

The influence of input variables on the model output is evaluated based on the correlation analysis. Correlation measures the strength and weakness of linear relationship between two variables. There are several coefficients that measure the correlation degree and Pearson's correlation coefficient is used to determine the input variables relevance for this paper. Pearson's correlation coefficient is calculated by dividing covariance of two variables by product of their standard deviation as shown in equation (A.1 – A.2), where $r$ represents Pearson's correlation coefficient. In equations (A.1-A.2), $\text{cov}(xy)$ is covariance which represents strength of linear relationship between two variables $x$ and $y$; $\bar{x}$ and $\bar{y}$ are mean values of variables $x$ and $y$; $s_x$ and $s_y$ are standard deviations of variables $x$ and $y$; and $n$ is the number of data.

$$r = \frac{\text{cov}(xy)}{s_x s_y} \tag{A.1}$$

$$\text{cov}(xy) = \frac{1}{n-1} \sum_{i=1}^{n} (x_i - \bar{x})(y_i - \bar{y}) \tag{A.2}$$

The correlation coefficients can range from -1 to +1:

$r = 1$ : perfect positive linear correlation

$r = -1$ : perfect negative linear correlation

$0.1 < |r| < 0.25$ : small positive linear correlation

$0.25 < |r| < 0.6$ : medium positive linear correlation

$0.6 < |r| < 1$ : strong positive linear correlation

$-1 < r < 0$ : negative linear correlation

Climatic conditions (outside temperature and solar radiation), operational power level characteristics and approximate occupancy profile are used to evaluate the relevance variables that affect building heat demand based on case study data. Other variables pseudo dynamic transitional attributes, which signifies the dynamics of building characteristics is not consider for relevance variable determination since it only signifies time and phase interval of heating power transition.

Results show the linear coefficient of correlation of outside air temperature, solar radiations, occupancy profile and operational power level characteristics with the heat load are -0.84, -0.40, 0.32 and 0.35 respectively. Results, thus, signifies that climatic conditions (outside temperature and solar radiations) are relevant input variables to predict the heat load. Also, it is clearer that occupancy profile and operational power level characteristics has medium positive correlation with heat load and shows relevance to characterize the heat demand behaviour.